# Application of Color Block Code in Image Scaling

Hao Wang
*Network and Information
Technology Centre*
*Beijing University of Technology*
Beijing, China
wanghao@bjut.edu.cn

Yu Bai
*Medical Engineering Division*
*Beijing Friendship Hospital*
Beijing, China
blancaby@sina.com

Jie Liu
*Information Department*
*Beijing University of Technology*
Beijing, China
liujie217@bjut.edu.cn

Guangmin Sun
*Information Department*
*Beijing University of Technology*
Beijing, China
gmsun@bjut.edu.cn

Yanjun Zhang
*Network and Information
Technology Centre*
*Beijing University of Technology*
Beijing, China
zhangyj@bjut.edu.cn

Jie Li
*Information Department*
*Beijing University of Technology*
Beijing, China
lijie00802@163.com

*Abstract*—**Aiming at the high cost of embedding annotation watermark in a narrow small area and the information distortion caused by the change of annotation watermark image resolution, this paper proposes a color block code technology, which uses location information and color code to form recognizable graphics, which can not only simplify the annotation graphics, but also ensure the recognition efficiency. First, the constituent elements of color block code are designed, and then the coding and decoding method of color block code is proposed. Experiments show that color block code has high anti-scaling and anti-interference, and can be widely used in the labeling of small object surface and low resolution image.**

## I. INTRODUCTION

Adding annotation information to image RGB sequence is an important means of image archiving, traceability and anti-counterfeiting. However, in the process of practical application, due to the influence of the main content of the image, it is difficult to mass produce the marking code or marking watermark. For example, after marking the image with QR code, the image copy will be subject to scale transformation, clipping, down sampling and other operations, which will affect the integrity and readability of the marking content, such as information matrix QR code, Chinese sensible code, PDF417 code is widely used in the fields of Internet, newspapers, document management and so on. The essence of these methods is to store the data in recognizable graphics in binary encoding, and the scanning equipment can quickly identify the encoded content. The content of such graphics is complex. After generating a fixed resolution image, if the image is scaled or down sampled with high intensity, the code is easy to be changed or lost. If the recognizable QR code is re-attached to the image with reduced resolution, its main components may be covered and the availability of thumbnails may be reduced too.

Labeling microfluidic chips usually requires the help of expensive laser equipment. It is envisaged that the reagent material is mixed into ink and the color block code(CB code) is printed to the paper-based microfluidic chip to be marked with an inkjet printer [1], which can not only avoid the coding deviation caused by particle agglomeration, but also effectively reduce the marking cost.

The CB code method proposed in this paper is suitable for image labeling and recognition in a more limited space. This method can effectively make up for the shortcomings of low data volume and narrow coding range of bar code. It is superior to QR code in anti-scaling distortion and additional image stability. Through the form of "live code", the color matrix can make the annotations have the advantages of high density, large capacity, good confidentiality, low cost and easy production. It can be used for precision instrument labeling, ultra-low resolution image labeling, and can also be used as data labeling in microfluidic devices. The CB code is composed of at least 16 pixels or color blocks, with obvious data representation method and high recognition rate. The labeling process is simple and fast, does not affect the characteristics of the marked subject, improves the labeling efficiency and reduces the cost.

## II. APPLICATION STATUS

Image archiving is more objective than text archiving, and plays an irreplaceable role in medical, remote sensing, exploration and other industries. Sometimes only the image file cannot completely describe the archiving attributes, so it is necessary to add description attributes or tags on the image, so that more comprehensive information can be obtained in the data application. However, the biggest problem of the image is that its marking and management are difficult to be automated and batch, because at present, the best method is to embed "annotation watermark" in the image to mark the image data content. Some authors use "fragile watermark" to protect the integrity of image data [2], and use "robust watermark" to protect image copyright [3], but when embedding watermarks in batches, marking the size and position of watermarks may damage the main features of the image itself. Fragile watermarks and robust watermarks need to use relatively complex algorithms to add and read [4]. None of the schemes are very applicable. In the image application stage, in order to improve the speed of image processing, the original image is often compressed or scaled, and the existing annotation information will also be lost or wrong. The marked content needs encryption processing or authorized browsing, so the marked content should also have a certain access control function.

For example, when embedding the index watermark into the medical image file, the ethics committee needs to cut the image when authorizing and distributing the file, which requires that the watermark should be placed in a position that does not affect the main features and is not easy to lose. Another example is to embed the traceability watermark into the face photo. After being processed by the face recognition algorithm, a thumbnail with small resolution is generated and displayed on the recognizer screen. The image content has changed. As an image watermark, it needs to have scalable attributes. It should be clear after scaling, and the annotation information can still be read after generating a low-resolution copy or compression.

The CB code uses the low resolution color matrix as the annotation information to embed or replace the pixels of the host image, which can better solve the labeling problem of low resolution and high scaling image. It can not only maintain the image features, but also realize the recognition of annotation. This method is characterized by high coding efficiency, easy implementation and good recognition accuracy.

### III. ENCODING AND DECODING PRINCIPLE

The CB code proposed in this paper is a graphic matrix composed of $4 \times 4$ pixels color blocks, as shown in Fig. 1. Except that the positioning block (P-block) is represented by three fixed colors of red, green and blue, other blocks are arranged and combined by different colors. The higher the color contrast of the blocks, the higher the fault tolerance rate.

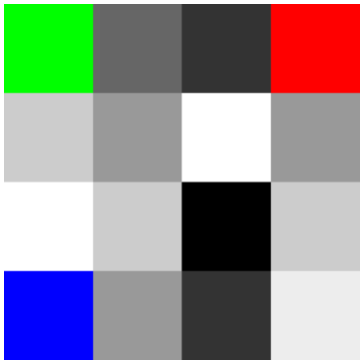

Fig. 1. CB code.

It can use at least 16 pixels as the coding area, which is enough to deal with the annotation of ultra-low resolution images. High contrast color blocks can maintain high recognition when carrying out large-scale transformation after labeling the host image, because image up-down sampling generally operates on neighborhood pixels, and the category of color can still be distinguished after CB code feature positioning, data sampling and color clustering.

### A. CB Code Composition

The CB code is composed of positioning block (P-block), data block (D-block) and verification block (V-block).

The P-block is located at the three vertices of the CB code, and counterclockwise is red (0xff0000), green (0x00ff00) and blue (0x0000ff).

D-block refers to all blocks except 4 vertices. The basic data consists of six color blocks: 0x000000, 0x333333, 0x66666, 0x999999, 0xcccccc and 0xFFFFFF.

The V-block is located at the fourth vertex different from the P-block, and surrounds the data area with the P-block. It is composed of three identical 8-bit color codes. The verification method can be Cyclic Redundancy Check (CRC) algorithm [5].

The color block can be a square image or a pixel. If the identification of pixels is affected by the performance of the camera, the method of identifying files can be selected.

### B. CB Code Positioning

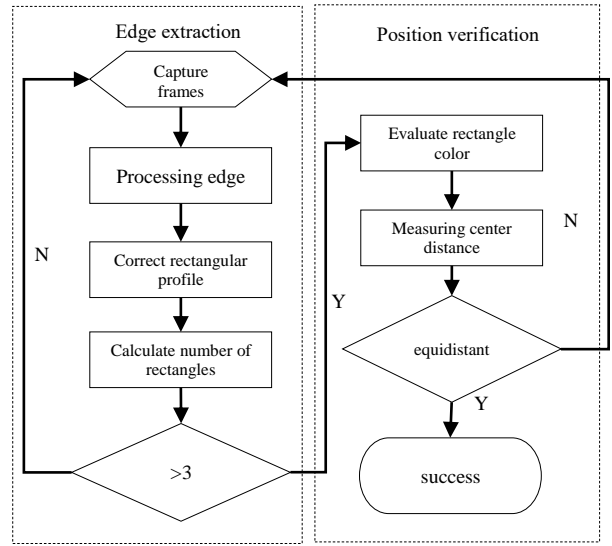

Fig. 2. Positioning flow chart.

The P-block is mainly responsible for guiding the program to lock the CB code area. The positioning process is divided into edge extraction [6], [7] and position verification, as shown in Fig. 2.

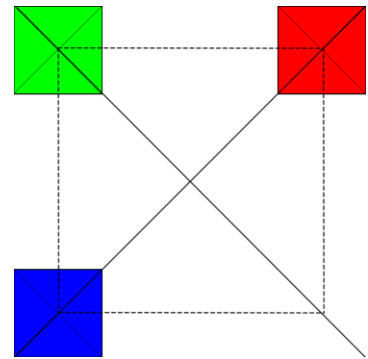

Fig. 3. Positioning sampling.

The color information of the current acquisition frame is segmented and morphologically processed to obtain the CB code candidate area. The Sobel algorithm and Hough transform are used to accurately extract the orthogonal grid in the area. The sampling points are processed according to the window proportion, the rectangular contour is corrected, and the number of rectangles is calculated to determine the CB code area. In order to calculate the color information of the location block more accurately, it is necessary to eliminate the background interference near the target, otherwise the interference of color block segmentation will affect the execution of subsequent target re detection [8]. Then carry out position verification, interlace and diagonal sampling for the colors of equidistant color blocks, and horizontally evaluate the identified color space, that is, complete the RGB evaluation according to the hue probability density classification. If the color blocks of three colors meet the RGB characteristics, and meet the equidistant conditions and color order, the positioning is successful. The positioning results are shown in Fig. 3.

*C. Color Initialization*

When CB code is recognized in the form of file, there is no need to initialize the color. The GM or GD library [9] can be directly used to identify the color code. The color image retrieval algorithm of Z search co-occurrence matrix can be used to retrieve the RGB pure color data [10], and return the location information and color information to provide data support for CB code positioning and data block recognition.

When collecting through the camera, due to the influence of illumination or white balance factors, it is necessary to normalize the color of the collected image and the target color, process the **RGB** spatial data of the collected image with linear gamma check, then map the **RGB** space to **XYZ** space [11], and then convert it into **XY** space respectively and normalize **Y**. the purpose of that is to make the brightness of the two images comparable, and Bradford transform [12] the two sets of data. The resulting matrix data is multiplied by the $D_{65}$ scale matrix [12] of **XYZ** space to obtain the coefficient matrix **M.** The **M** matrix is multiplied by the linear **RGB** matrix of the original image to complete color correction, and the image is saved through inverse linear conversion. The image is the CB code to be recognized after standard **RGB** verification, as shown in Fig. 4.

When the image collected by the camera is disturbed by electromechanical interference, the filtered Moire noise [13] around the point to be measured is detected by adaptive Gaussian filter. Thanks to the high peak signal-to-noise ratio of the pixel feature of the CB code, the periodic noise near the point to be measured can be filtered by using the automatic notch filter according to the detected Moire feature. In order to prevent the generated artifact from interfering with the positioning, after initializing the standard values of all colors in the CB code, if the maximum likelihood estimation value of the pixel to be tested is between the high and low bits of the standard value, it is set as the value of the adjacent pixel, otherwise the data remains unchanged. This method can also reduce the running time of clustering and classification in the data identification link.

*D. Data Identification*

Set the data color clusters as 0, 3, 6, 9, C and F respectively. These colors can ensure high contrast under the influence of low resolution or shot noise, which is conducive to improving the color recognition rate [14], [15]. The code clusters can provide at least $2.17 \times 10^9$ combinations. Users can adjust the color types of code clusters or color patches as needed. Ideally, it can produce up to $1.68 \times 10^{84}$ sets of data.

In order to improve the accuracy of D-block color clustering, the connected domain method and hole filling method can be used to process the data block after color initialization. The geometric features of color block arrangement and the linear segmentation area of positioning block can be used to match the data content and position. Set the confidence of data CB code cluster as *s*, as shown in Eq.1.

$$s(t_i, d_j) = \exp\left(-\left(\left|A_{t_i} - \delta A_{d_j}\right| + \left|AR_{t_i} - \delta AR_{d_j}\right|\right.\right.$$
$$+ \left|LWR_{t_i} - \delta LWR_{d_j}\right|$$
$$+ \left|L_{t_i} - \delta L_{d_j}\right| + \left|W_{t_i} - \delta W_{d_j}\right|$$
$$\left.\left.+ \frac{\|C_c - C_d\|}{2h_w n}\right)\right) \quad (1)$$

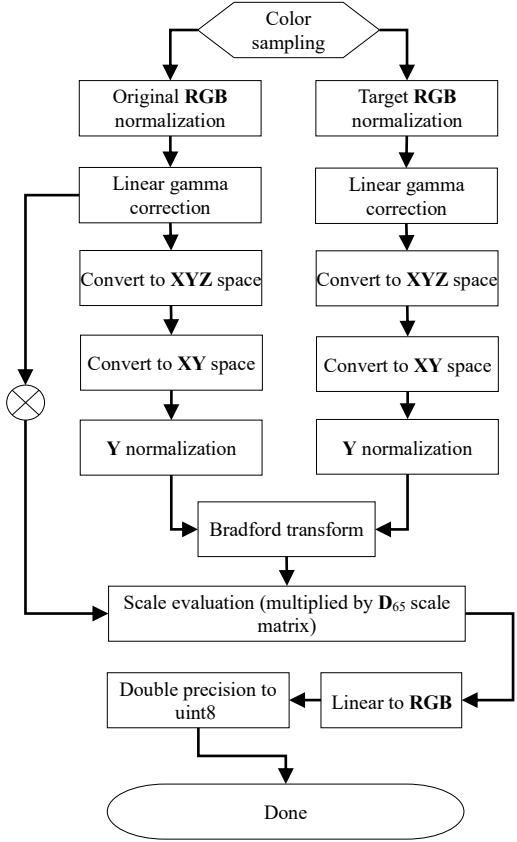

Fig. 4. Color initialization process.

Subscript *t* represents the color of the object code cluster, *d* represents the collected color value, and *A* is the size of the connected domain, $\delta$ is the scale factor. $AR$ represents the proportion of the current connected domain to the color block, *W* represents the color block width, *L* represents the color block length, *R* represents the color block length width ratio, and $C_c$ represents the center coordinate of the data color block, $C_d$ represents the coordinates of the collected color, $h_w$ is the longest edge of the collected color block. *n* represents the number of pixels discarded after the failure of matching the target code cluster. The smaller *n* is, the smaller re-detection range is. The confidence threshold can be adjusted manually. When *s* is greater than or equal to the threshold, clustering operation is carried out. Otherwise, it is discarded until the data block is detected. Finally, according to the arrangement order of D-blocks, a data set with color code cluster as the class name is formed, which is the data sequence of CB code.

The color code clustering method adopts the K-means++ method. This method is adopted because if the centroid is randomly distributed and the convergence is slow in traditional method, which affects the recognition cycle of CB code. The preset type of D-block is known, and the initialization of centroid into color code can quickly reduce the time-consuming of clustering process.

The coordinates of the two points in the three-dimensional Euclidean distance correspond to the three channels of the pixel respectively. The *k* value is the total number of colors preset in the D-blocks in CB code, $\mu_i$ is the centroid of the color code cluster $C_i$ and *x* is coordinate point. Eq.2 is satisfied.

$$\mu_i = \frac{1}{|C_i|} \sum_{x \in C_i} x \qquad (2)$$

Minimizing the square error *E* can make the pixel information read each time closer to the preset pixel code:

$$E = \sum_{i=1}^{k} \sum_{x \in C_i} \|x - \mu_i\|_2^2 \qquad (3)$$

This method is sensitive to noise and outliers, so the image should be filtered before clustering. Although the result is a local optimal solution, the value of the data in CB code cluster can also be accurately determined in combination with the weighted average value of the number of sampling points.

### E. Data Verification

The V-block is used to verify the correctness of the identified data sequence. If the CRC method is adopted, the polynomial *poly*=0x07 is used for verification, $G(x) = x^8 + x^2 + x + 1$ is used as the divisor. The D-block $F(x)$ performs a modulo-2 operation as a dividend, the remainder is used as the three channel data of the V-block. The remainder obtained after the operation of D-block $F(x)$ and polynomial $G(x)$ in Fig. 1 is 0xED, and the color of V-block is 0xEDEDED. Extend the remainder as check data to the end of data bit to obtain $F(x)'$. During verification, if the remainder is 0 after modulo-2 Division between $F(x)'$ and $G(x)$, the data verification is successful, otherwise it fails, it needs to be scanned and verified again.

## IV. EXPERIMENTS

### A. QR Code Scaling

Although the QR code does not rely on the network, and the covered blocks do not affect the recognition results of the equipment, the content covered by the graphics becomes complex with the increase of binary data, and the requirements for resolution are high. It can accommodate up to 2953 bytes [16], and there are few cases with a length of more than 150 characters in practical application, Because super long content and high fault tolerance will increase the coding density.

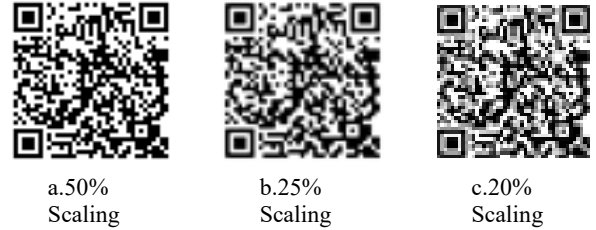

a.50% Scaling   b.25% Scaling   c.20% Scaling

Fig. 5. QR code image reduction effect.

If the generated QR code image is scaled or the resolution is reduced, recognition rate will be greatly reduced. As shown in Fig. 5, the resolution is $260 \times 260$ pixels, a QR code with a fault tolerance rate of 30% cannot be recognized after scaling by 20%. At present, the solution is to regenerate a new QR code and add it to the image again, or use the scaling method based on image principal component. The former has a large workload and loses the significance of image traceability, while the latter requires the equipment to bring its own scaling algorithm, which has a high cost. Experiments show that if the image resolution is lower than the QR code resolution or cannot avoid the main features, the scheme is completely ineffective.

### B. CB Code Positioning Experiments

The scanning program tests the CB code positioning status under the states of forward direction, rotation, distortion and color interference. A total of 100 rounds of grouping experiments show that the positioning is successful and "OK" is displayed. Otherwise, the results are not displayed, as shown in Fig. 6.

The scanning program can divide the CB code into 16 rectangular areas and identify 3 positioning codes. The positioning success rate is more than 99% and the positioning time is less than 1000 milliseconds. The samples are displayed on the LED screen, indicating that the positioning stage is not disturbed by Moire and shot noise.

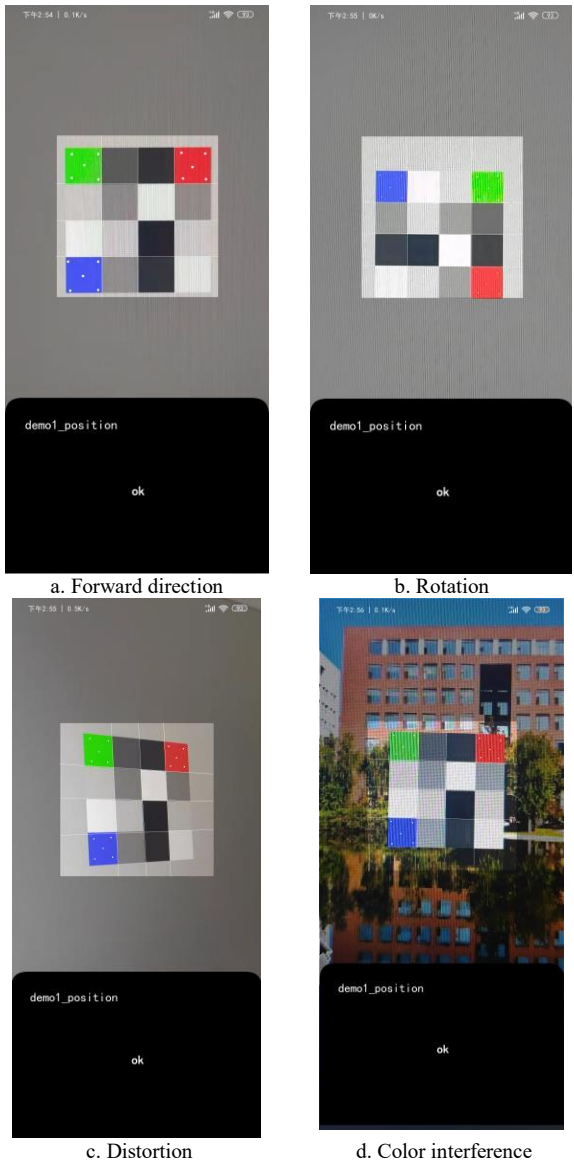

a. Forward direction      b. Rotation

c. Distortion      d. Color interference

Fig. 6. Positioning experiment.

*C. CB Code Data Recognition Experiment*

- File identification

Upload the CB code picture to the identification program to accurately obtain the data and CRC code. The file recognition program is based on B/S architecture. Python is responsible for edge detection and Hough transform. The PHP script is responsible for image file filtering, compression, marking, color recognition and data feedback. Because the image file carries accurate and rich pixel information, the data block color information is less disturbed by noise, and the recognition rate can be maintained at more than 99.9%, as shown in Fig. 7.

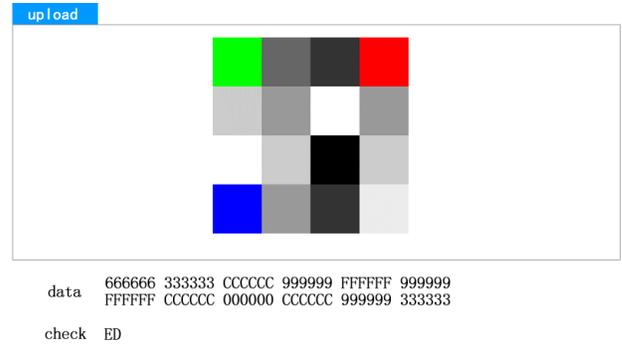

Fig. 7. Online CB code system based on PHP.

Although the accuracy of CB code recognition in the form of file is high, there are also special cases. For example, if the CB code is attached to an image with similar characteristics, as shown in Fig. 8, the recognition rate will be reduced or even unable to locate, because the positioning and edge detection algorithm mistakenly recognizes the grid of the background picture as the edge of the CB code, and the positioning code feature is not retrieved in the grid. The solution is to isolate the background image and CB code image with a solid color frame, as shown in Fig. 9, or add the recognition program to the function of manually selecting the area to be inspected.

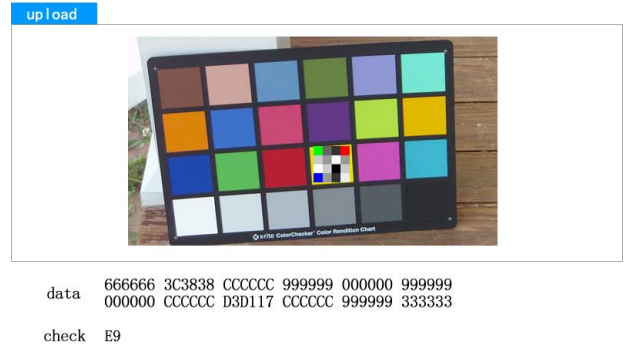

Fig. 8. CB code in complex images.

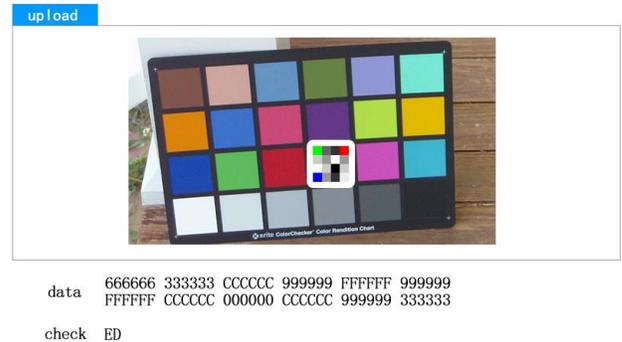

Fig. 9. CB code with border in complex images.

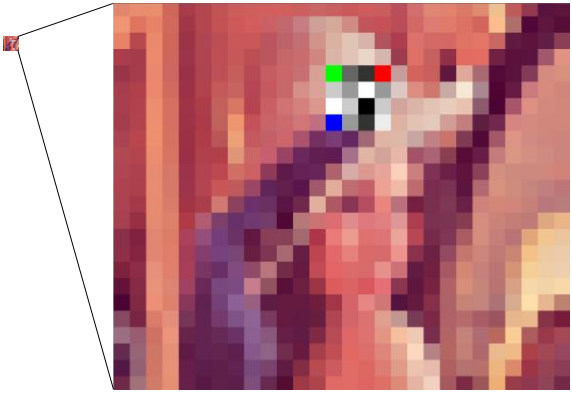

Fig. 10. CB code marking Lena 28 ×28 image.

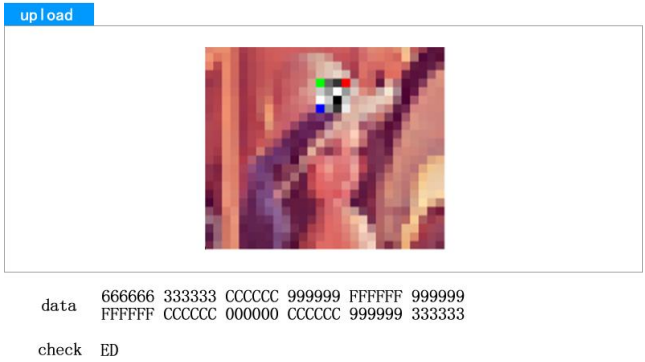

Fig. 11. Calculating of 28 ×28 Lena image with CB code.

Reduce the resolution of Lena photos to 28 × 28 pixels. In this state, the annotation of QR code cannot be applied, and the CB code can be accurately marked on her hat without affecting the face and other main parts of the image, as shown in Fig. 10. Upload the image to the online recognition program, and the accuracy of CB code recognition is greater than 99%, as shown in Fig. 11, which shows that the recognition of CB code is independent of image size.

- Camera identification

When using the camera to identify the printed CB code images(reflective identification), without the influence of serious uneven light and low light, the average time is only 500ms, and the accuracy can reach 99.9%. The CB code (self-luminous) on the screen needs to deal with Moire interference. The minimum time consumption is 1200ms, and the correct recognition rate is more than 70%. The overall time-consuming of the recognition process is directly related to the specification of the photosensitive element of the equipment and the running speed of the processor. The higher the specification of the photosensitive element of the camera, the smaller the error. The performance of the processor is positively related to the speed of positioning and verification. As shown in Table I, in the recognition experiment when the CB code is 260 × 260 pixels resolution, the equipment accuracy of 4mp in the eighth group of experimental results is 0, which does not mean that it cannot be recognized, because the program execution process exceeds the limited time. In order to optimize the user experience and prevent the recognition program from faking death in the V-block color recognition link, the timeout is set to 6000ms. If the verification is not successful within this time, the next frame will be rescanned.

TABLE I. STATISTICAL OF CB CODE RECOGNITION ON SCREEN

| Group | Resolution (MP) / refresh rate (Hz) | Average accuracy (%) | Average time (ms) |
|---|---|---|---|
| 1 | 48/50 | 70.5 | 1300 |
| 2 | 48/60 | 71.5 | 1200 |
| 3 | 12/50 | 74.8 | 1450 |
| 4 | 12/60 | 79.1 | 1450 |
| 5 | 8/50 | 70.1 | 1330 |
| 6 | 8/60 | 70.4 | 1300 |
| 7 | 6/50 | 40% | 6000 |
| 8 | 4/50 | 0% | -- |

*D. Occlusion Experiment*

In order to verify the robustness of the color block code, the experiments of area occlusion, color occlusion and reflection are carried out respectively, as shown in Fig. 12.

Block the 7th, 8th, 9th, 11th and 12th blocks and their boundaries in Fig. 12a with words, and the color information can be recognized. The recognition accuracy is directly proportional to the number of color sampling points [17]. The experimental results are consistent with those without blocking.

If the block unit is 100% covered with interference color, as shown in Fig. 12b, the 3rd color block completely loses the characteristics of color code cluster and cannot be recognized.

The color belonging to code cluster C is used for interference, as shown in Fig. 12c. Although the block code can be located, the interference color has a 50% probability of being used as the color of the D-block to participate in the operation, resulting in verification failure and unable to output the result.

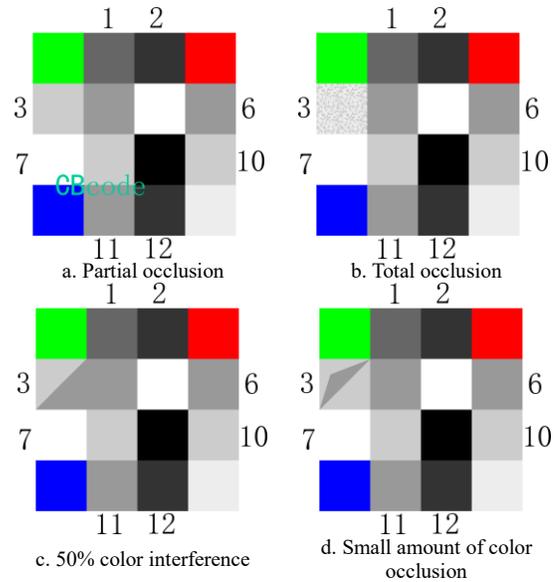

Fig. 12. CB code interference experiment.

When the interference color coverage is less than 50% of the block area, as shown in Fig. 12d, it can be correctly identified,

the accuracy and speed are the same as those without color coverage. This experiment shows that the CB code can resist occlusion attacks less than 50%.

When the CB code on the printed or non-self-luminous medium is affected by the reflected light, there is no sudden change in the color of the D-block and the V-block, it usually only affects the recognition time. When there is no output result, the equipment angle can be adjusted. 200 rounds of random interference tests are carried out on the same CB code. When the coverage of color block unit is less than 50%, the fault-tolerant rate is 31%, which is equivalent to that of QR code(matrix).

If the V-block and the D-block produce errors at the same time, and the error data exactly conforms to the calibration formula, the error result is output, for example, the V-block that should have been 0xED is identified as 0xE8, the 12th bit data should be 0x333333, which is identified as 0x666666. The remainder of this group of values is 0 after polynomial operation verification, so the algorithm mistakenly believes that the identification is successful and outputs the wrong result. The solution is to reduce the colors of D-block or improve the contrast of D-block.

*E. Scale Transformation Experiment*

When the QR code is scaled to 20%, it cannot be recognized in the form of file or image. The interpolated samples of CB code with 260 ×260 pixels resolution are shown in Fig. 13a-d. the data can be read normally, the recognition accuracy is more than 69%, and the time-consuming is less than 6000ms. The recognition accuracy of image scaling rate between 12.5% and 50% is more than 70%, and the time-consuming is kept below 2000ms, as shown in Fig. 14. The image down sampling in Fig. 13e causes too much missing content, and the image content cannot be recognized after up sampling. Fig. 13f shows the CB code with 4 ×4 pixels resolution. After scaling 6400%, the data is clearly visible and the image can be fully recognized.

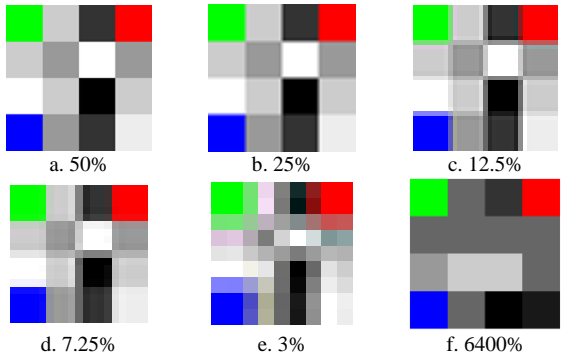

| a. 50% | b. 25% | c. 12.5% |
| d. 7.25% | e. 3% | f. 6400% |

Fig. 13. CB code scaling experiment.

The recognition efficiency is affected by the number of sampling points. The more sampling points, the higher the accuracy of identifying a single color block and the longer the time-consuming. However, too low a color block sampling point will also increase the recognition time and may reduce the accuracy.

If the V-block and code cluster classification are not accurate enough, the verification cannot be passed, and the whole process will cycle all the time.

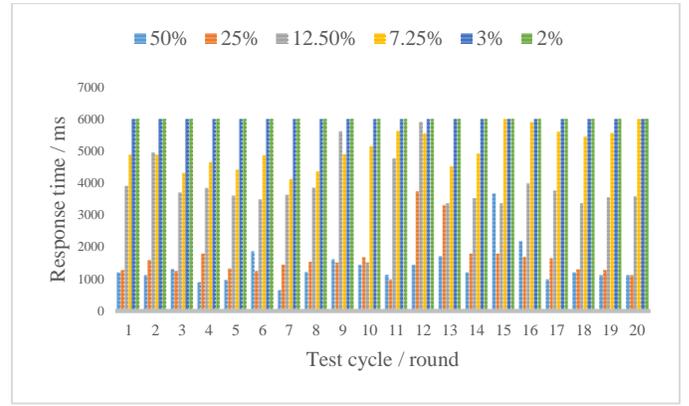

Fig. 14. CB codes response experiments.

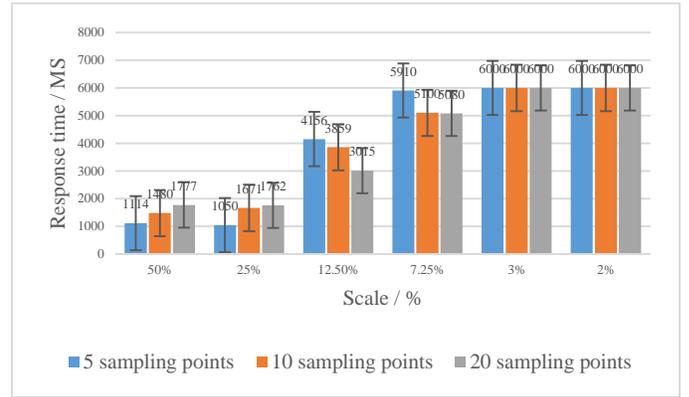

Fig. 15. Sampling contrast experiment.

Sample 5, 10 and 20 pixels respectively for color blocks of different specifications. The optimal number of sampling points is 5 when the size is reduced by 50% to 12.5%, and the efficiency is the highest when the sampling point of CB code is 20 when the size is reduced by 12.5% to 7.25%, as shown in Fig. 15.

When the CB code is scaled, the sampling coefficient is generally less than 1, which will inevitably lead to the lack of pixel position information and signal aliasing.

TABLE II. COLOR RECOGNITION ERROR TABLE

| round | Block 1 | Block 2 | Block 3 | Block 4 | Block 5 |
|---|---|---|---|---|---|
| 1 | D4C9D4 | 303535 | CBCBCB | 979897 | FFFFFF |
| 2 | CECDCE | 303131 | C9CAC9 | 919191 | FFFFFF |
| 3 | CACACA | 2C2C2C | CFCECF | 979797 | FFFFFF |
| 4 | CBC5CB | 253434 | C1C1C1 | 929292 | FFFFFF |
| 5 | CFCACF | 3F3E3E | C8C8C8 | 909090 | FFFFFF |
| 6 | DADADA | 323333 | CFCCCF | 989898 | FFFFFF |
| 7 | CECACE | 283636 | D0CDD0 | 979797 | FFFFFF |
| 8 | D2CDD2 | 422F2F | C8C8C8 | 979797 | FFFFFF |
| 9 | CFCACF | 293636 | DCCCDC | 939393 | FFFFFF |
| 10 | D8D8D8 | 323434 | D1CBD1 | 969696 | E8E8E8 |

If the scaling rate is too small, almost all colors except red, green, blue, black and white will be distorted. Table II shows some data of 1 to 5 blocks after scaling the CB code of 260 × 260 pixels resolution to 7.25%. The error data can be corresponding to the code cluster, as shown in Fig. 16. Through classification, the data are 0xCCCCCC, 0x333333, 0xCCCCCC,

0x999999 and 0xFFFFFF respectively, Exactly matches the preset color.

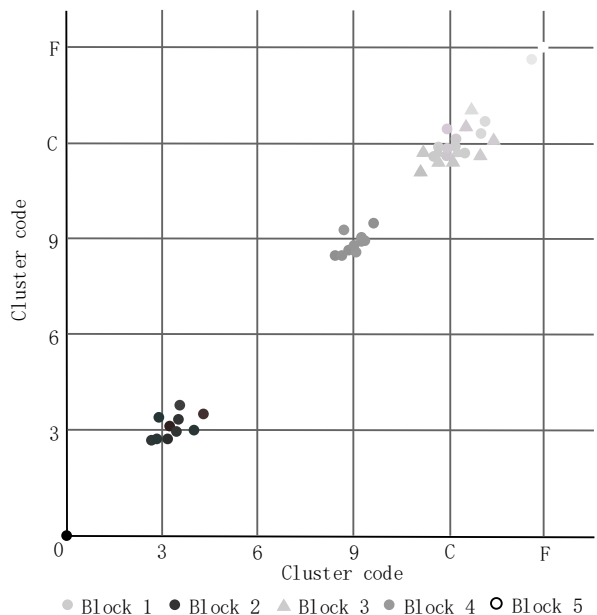

Fig. 16. Color clustering experiment.

## V. CONCLUSION

CB code can effectively make up for the shortcomings of complex and easy distortion such as QR code, and has the characteristics of high efficiency, low cost and high availability.

CB code is applicable to ultra-low resolution images, surface printing of small objects, large-scale scale transformation, etc., with a minimum resolution of $4 \times 4$ pixels, the influence of CB code composed of a single pixel on the main features of the image can be ignored.

The color block unit has good anti-interference performance when the coverage is less than 50%, and the recognition rate remains above 69%.

The data capacity of CB code can adapt to general scene applications. Live code is used to record data, with flexible coding mode and high security.

The recognition speed of CB code is affected by the equipment performance, and the recognition speed will continue to be improved in the follow-up work.

Complexity of image content and ambient light is directly related to the recognition accuracy of CB code. It may be necessary to use neural network model to define the color category of D-block. This problem will be studied later.

The printing of paper-based microfluidic chip is feasible in theory, but the printing effect cannot be verified due to the influence of experimental environment. This experiment will be completed in the follow-up work.